\documentclass[aps,prl,twocolumn,superscriptaddress,showpacs=52.38.Kd]{revtex4}

\usepackage{graphicx}

\begin{document}

\title{Beam loading in the nonlinear regime of plasma-based acceleration}

\author{M. Tzoufras }
\affiliation{Department of  Electrical Engineering, University of California, Los Angeles, CA 90095}
\author{W. Lu }
\affiliation{Department of  Electrical Engineering, University of California, Los Angeles, CA 90095}
\author{F. S. Tsung }
\affiliation{Department of Physics and Astronomy, University of California, Los Angeles, CA 90095}
\author{C. Huang}
\affiliation{Department of Physics and Astronomy, University of California, Los Angeles, CA 90095}
\author{W. B. Mori }
\affiliation{Department of  Electrical Engineering, University of California, Los Angeles, CA 90095}
\affiliation{Department of  Physics and Astronomy, University of California, Los Angeles, CA 90095}
\author{T. Katsouleas }
\affiliation{Department of Electrical Engineering, University of Southern California, CA 90089}
\author{J. Vieira }
\affiliation{GoLP/Instituto de Plasmas e Fus\~{a}o Nuclear,  Instituto Superior T\'{e}cnico, 1049-001 Lisboa, Portugal}
\author{R. A. Fonseca }
\affiliation{GoLP/Instituto de Plasmas e Fus\~{a}o Nuclear,  Instituto Superior T\'{e}cnico, 1049-001 Lisboa, Portugal}
\author{L. O. Silva }
\affiliation{GoLP/Instituto de Plasmas e Fus\~{a}o Nuclear,  Instituto Superior T\'{e}cnico, 1049-001 Lisboa, Portugal}

\today

\begin{abstract}
A theory  that describes how to load negative charge into a nonlinear, three-dimensional plasma wakefield is presented. In this regime, a laser or an electron beam blows out the plasma electrons and creates a nearly spherical ion channel, which is modified by the presence of the beam load. 
Analytical solutions for the fields and the shape of the ion channel are derived. It   is shown that  very high beam-loading efficiency can be achieved, while the energy spread of the bunch is conserved.  The theoretical results are verified with the Particle-In-Cell code OSIRIS. 
\end{abstract}

\maketitle

Plasma-based acceleration relies on an underdense plasma to transfer the energy from a  laser beam or an electron beam to a trailing bunch of electrons or positrons \cite{Dawson,pchen}. The beam load is  accelerated in a wake moving with a velocity near the speed of light, $c$, 
 until the driver's energy is exhausted or---in the case of a laser driver---until it outruns the plasma wave.   
Major advances in both laser- and beam-driven accelerators have recently been achieved in a regime in which the fields of the driver are so intense, they push all plasma electrons aside, generating a pure ion channel \cite{RosenzWEIg}.

For the laser-driven accelerator \cite{Dawson}, experiments \cite{Mangles,leemans} have inferred and simulations \cite{Pukhov,tsung,tsupop} have demonstrated that electrons self-injected into the ion channel can form a quasi-monoenergetic  beam.  Externally injected, low-charge bunches have been shown to improve the reproducibility and the quality of the final electron beam \cite{malka}. In the beam-driven case  \cite{pchen},  stable acceleration of the tail of a $42GeV$ electron beam culminated in the doubling of the energy of some electrons in less than $1$ meter \cite{miaomiao}. 
  The high-gradient acceleration of a short trailing electron bunch in the wake of a driving beam, which is central to the afterburner concept \cite{afterburner},  has  also been achieved \cite{themos}.  
A theory that describes the wakefield in this blowout regime has recently been  developed \cite{WEI,weipop}.

While there has been tremendous progress both experimentally and theoretically on understanding how wakes are excited in the 3D nonlinear 
 regime,  there has been little work on how  the trailing beam loads the wake. In the linear regime, the issue of beam loading  was addressed in Ref.  \cite{Katsouleas}, where the wakefield generated by the trailing bunch was superimposed on that of 
 the driver to yield the final accelerating field.  Thus, the maximum charge that can be loaded was evaluated and   the current profile that makes the wakefield within the bunch flat was determined. 
Ref.  \cite{Katsouleas} also discussed the  effects of transverse beam loading, emittance, and phase slippage, and concluded that  beams with spot sizes much smaller than those of the  wake are required.

 When blowout occurs, the accelerating field is identical within each transverse slice of the ion channel \cite{WEI,weipop}, so as opposed to the linear regime, transverse beam loading does not affect the energy spread of the beam.  Additionally, because the focusing force in the ion channel is linear, the emittance is conserved. Therefore, the most important consideration for reducing the energy spread is keeping the accelerating field constant along the propagation direction. We note that  for high-energy physics applications,   
  narrow trailing bunches are still needed for matched beams and for reducing synchrotron radiation losses \cite{Katsouleas}.

In an estimate offered in Ref.  \cite{weiprstab},  the number of particles was found to  scale with the normalized volume of the bubble (or the square root of  the laser power). The same scaling was  obtained in Ref. \cite{Gordienko} but the coefficient, determined by simulations, was more than three times larger than the one estimated in Ref. \cite{weiprstab}. These results   
 are not necessarily contradictory because, in principle, one can  choose to accelerate either a small number of particles to high energy or a large number of particles to low energy. The question of merit is not just how many electrons can be loaded, but  what kind of electron bunch can most efficiently convert  the energy available in the wake of the driver into kinetic energy uniformly distributed to its electrons.

\begin{figure}[!thb]
\begin{center}
\includegraphics[width=3.2 in]{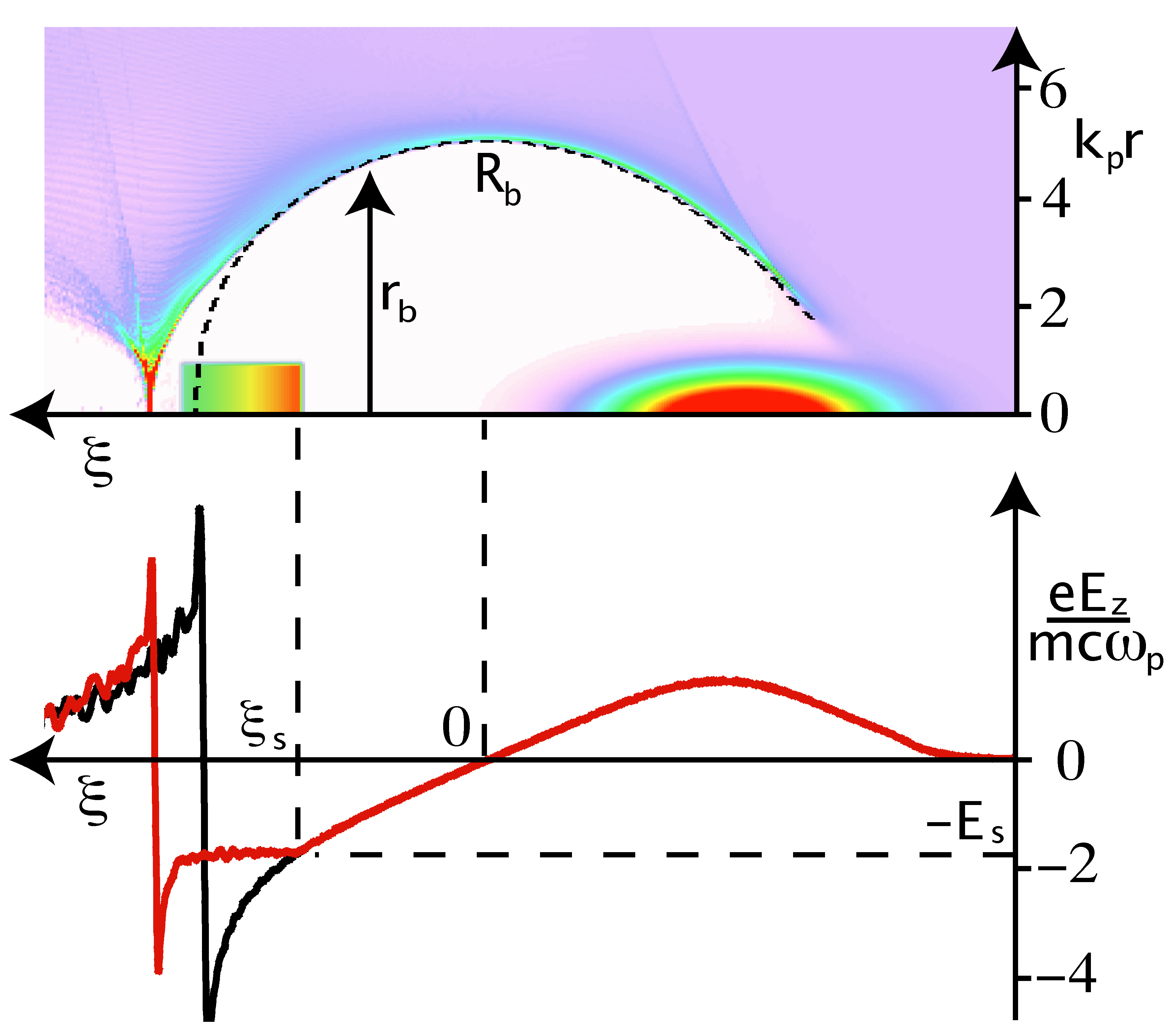}
\caption{  
 The electron density  
from a PIC simulation with OSIRIS \cite{Fonseca} 
 for $k_pR_b=5$ is presented. The  beams 
  move to the right. The broken black line traces the blowout radius 
   in the absence of the load.  On the bottom, the red (black) line is the lineout of the wakefield $E_z(\xi,r_b=0)$ when the beam load is present (absent).} \label{figure1}
\end{center}
\end{figure}

 In Ref. \cite{WEI}, the wakefield $E_z$ in each transverse slice was found to be proportional to the product of the local radius of the ion channel $r_b$  and the slope $dr_b/d\xi$, where $(\xi,r_b)$ are cylindrical coordinates with $\xi=ct-z$ and the driver is moving toward positive $z$.  The shape of the bubble is represented by the trajectory of the innermost particle given by  Eq. ($11$) in Ref. \cite{WEI}.    This description  is valid between the points where the particle trajectories cross,  at the very front and the very back of the bubble.  To make progress analytically, we take  the ultrarelativistic limit, where the normalized maximum 
  radius of the ion channel is $\omega_pR_b/c\gg 1$. The equation for the  innermost particle trajectory reduces to (see Ref. \cite{WEI}):
\begin{equation}\label{differ}
r_b\frac{d^2r_b}{d\xi^2}+2\biggl[\frac{dr_b}{d\xi}\biggr]^2+1 = \frac{4\lambda(\xi)}{r_b^2}
\end{equation}
where we adopt normalized units, with length normalized to the skin-depth $c/\omega_p$, density to the plasma density $n_p$, charge to the electron charge $e$, and fields to $mc\omega_p/e$.  
The term on the right hand side of  
Eq. (\ref{differ})  can describe
 the charge per unit length of an electron beam and/or the ponderomotive force of a laser \cite{WEI}. Here we are interested in the back half of the bubble, where the wakefield is accelerating and the quantity $2\pi\lambda(\xi)$, with $\lambda(\xi) = \int_0^{\infty}rn_bdr$, is the charge per unit length of the beam load.

We define $\xi=0$ at the location where $r_b$ is 
 maximum, i.e.,    $\frac{dr_b}{d\xi}\vert_{\xi=0}=0$. In Ref. \cite{WEI}, it was shown that for $\omega_pR_b/c\gg1$, the wakefield is $E_z\simeq\frac{1}{2}r_b \frac{dr_b}{d\xi}$; therefore, $E_z(\xi=0)\simeq0$. For $\xi>0$, the electrons are attracted by the ion channel back toward the $\xi$-axis with $\frac{dr_b}{d\xi}\vert_{\xi>0}<0$ until $\xi=\xi_s$ where beam loading starts. For $\xi\geq\xi_s$, the electrons feel the repelling force from the charge of the accelerating beam, 
   in addition to the force from the ion channel. The additional repelling force decreases the slope of the sheath $\frac{dr_b}{d\xi}$, thereby lowering the magnitude of $E_z$. 
This can be seen in the simulation results in Fig. \ref{figure1}, where the trajectory of the innermost electron for 
 an unloaded wake is drawn on top of the electron density for a loaded wake, and the corresponding 
wakefield for the two cases is also plotted. The method for choosing the charge profile of the load is described below. 

 If the repelling force   
is too large and the beam too long,  the electrons in the sheath will reverse the direction of their transverse 
velocity at some $\xi_r$, where $\frac{dr_b}{d\xi}\vert_{\xi=\xi_r}=0$, and consequently $E_z(\xi_r)=0$. This is a very undesirable configuration because it implies that the front of the bunch feels a much stronger accelerating force than the back. 

We are interested in trajectories for which $r_b(\xi>0)$ decreases monotonically. $\lambda$ may then be expressed as a function of $r_b$: $\lambda(\xi)=l(r_b)$.  
 Substituting $r_b^{\prime\prime}=r_b^\prime\frac{dr_b^\prime}{dr_b}$, where the prime denotes differentiation with respect to $\xi$, 
Eq. (\ref{differ}) reduces to  $\frac{dr_b^\prime}{dr_b}=\frac{4l(r_b)-r_b^2[2(r_b^\prime)^2+1]}{r_b^3r_b^\prime}$, which can be integrated to yield: 
\begin{equation}\label{genwake}
E_z\simeq\frac{1}{2}r_b \frac{dr_b}{d\xi}=-\frac{r_b}{2\sqrt{2}} \sqrt{\frac{16\int^{r_b}l(\zeta) \zeta d\zeta+C}{r_b^4}-1}
\end{equation}

Before further analyzing 
  Eq. (\ref{genwake}), we comment on salient features of the unloaded case $(l(r_b)=0)$. Evaluating the constant in  Eq. (\ref{genwake}) from 
the condition $E_z(r_b=R_b)=0$, we obtain:
\begin{equation}\label{nochaw}
E_z(r_b)\simeq\frac{1}{2}r_b \frac{dr_b}{d\xi}=-\frac{r_b}{2\sqrt{2}} \sqrt{\frac{R_b^4}{r_b^4}-1}, R_b\geq r_b>0
\end{equation}
Eq. (\ref{nochaw})  can be integrated from the top of the bubble $r_b(\xi=0)=R_b$ to yield the innermost particle trajectory for $0< r_b\leq R_b$:
\begin{equation}\label{empty_bubble}
\frac{\xi}{R_b}=
2E\biggl(\arccos\Bigl(\frac{r_b}{R_b}\Bigr)\bigg\vert\frac{1}{2}\biggr)
-F\biggl(\arccos\Bigl(\frac{r_b}{R_b}\Bigr)\bigg\vert\frac{1}{2}\biggr)
\end{equation}
where $F(\varphi\vert m),E(\varphi\vert m)$ are the incomplete elliptic integrals of the first and second kind  \cite{elliptic}. 

To minimize the energy spread on the beam, we seek the beam profile that results in $E_z(r_b\leq r_s)=\frac{1}{2}r_b\frac{dr_b}{d\xi}\vert_{r_b=r_s}
\simeq \mbox{constant}\equiv - E_s$ within the bunch. The shape of the bubble 
in this case is described by the parabola $r_b^2=r_s^2-4E_s (\xi-\xi_s)$.   For $0\leq\xi\leq\xi_s$, $E_z$ is given by Eq. (\ref{nochaw}). $E_s$ is found by requiring that the wakefield is continuous at  $\xi_s$: $E_s=\frac{r_s}{2\sqrt{2}} \sqrt{\frac{R_b^4}{r_s^4}-1}$. For $\xi_s\leq\xi\leq\xi_s+\frac{r_s^2}{4E_s}$, where $\xi_s+\frac{r_s^2}{4E_s}$ is the location at which the sheath reaches the $\xi$-axis, the profile of $\lambda(\xi)$ that leads to a constant wakefield is trapezoidal with maximum at $\lambda(\xi_s)=\sqrt{E_s^4+\frac{R_b^4}{2^4}}$ and minimum at $\lambda(\xi_s+\frac{r_s^2}{4E_s})=E_s^2$:
\begin{equation}\label{trapb}
\lambda(\xi)=\sqrt{E_s^4+\frac{R_b^4}{2^4}}-E_s(\xi-\xi_s)
\end{equation}
and he total charge $Q_s=2\pi\int_{\xi_s}^{r_s^2/(4E_s)}\lambda(\xi)d\xi$ is:
\begin{equation}\label{prbal}
Q_s\times E_s=\frac{\pi R_b^4}{16}
\end{equation}
Eq. (\ref{prbal}) illustrates the tradeoff between the number of particles that can be accelerated and the accelerating gradient. Because $Q_s\times E_s$ is the energy absorbed per unit length,  
Eq. (\ref{prbal})  also indicates that the efficiency from the wake to the beam load does not depend on the field $E_s$, as long as the charge profile is chosen appropriately. In Fig. \ref{figure1}, the charge of the  beam load was chosen using Eq. (\ref{prbal}) for $E_s=0.35R_b$. For this $E_s$, the location $\xi_s$ can be  obtained either from a simulation for an unloaded wake or from Eq. (\ref{nochaw})-(\ref{empty_bubble}), and the charge profile from Eq. (\ref{trapb}).

In a linear wake, a wide electron bunch with total charge $Q_l$ and  transverse spot size $A$, loaded at  some $\xi_0$  with the appropriate profile  \cite{Katsouleas} can also lead to a  flat  wakefield ($E_l=\frac{m\omega_pc}{e}\frac{n_1}{n_0}\cos(k_p\xi_0)$) within the bunch. The total accelerating force is: $Q_l\times E_l =\frac{E_0^2}{8\pi} A (1-\frac{E_l^2}{E_0^2})$, where $\frac{E_0^2}{8\pi} A$ is the energy per unit length of the wake in front of the bunch, and $\frac{E_l^2}{8\pi}A$ that behind it. The efficiency, $(1-\frac{E_l^2}{E_0^2})$,  increases for a decreasing accelerating gradient and reaches $100\%$ for $E_l=0$.  
This is in stark contrast  
to the blowout regime, where the efficiency $\eta_b$ is constant for any $E_s$.

To calculate the efficiency $\eta_b$, let us assume that  the bunch is terminated at  some $\xi_f$, where $\Delta\xi_f\equiv \xi_f-\xi_s<\frac{r_s^2}{4E_s}$. After this point the wakefield is described by Eq. (\ref{genwake}) with $l(r_b\leq r_f\equiv r_b(\xi_f))=0$.  From the boundary condition $E_z(r_f)=-E_s$, we obtain  $E_z(0<r_b\leq r_f)\simeq-\frac{r_b}{2\sqrt{2}} \sqrt{\frac{\widetilde{R}_b^4}{r_b^4}-1}$, where $
\widetilde{R}_b^4=R_b^4\biggl(\frac{r_f^4}{R_b^4}+\frac{r_f^2}{r_s^2}-\frac{r_f^2r_s^2}{R_b^4}\biggr)
$. In  Ref. \cite{weiprstab}, it was shown that the energy in the fields of a bubble is proportional to $R_b^5$. Therefore, the energy per unit length available to the bunch $\mathcal{E}_{\mbox{avail}}$ scales as $R_b^4$, and that left behind it $\mathcal{E}_{\mbox{lost}}$ scales as $\widetilde{R}_b^4$. We define the beam-loading efficiency as $\eta_b\equiv (\mathcal{E}_{\mbox{avail}}-\mathcal{E}_{\mbox{lost}})/\mathcal{E}_{\mbox{avail}}$. 
 Substituting the expression for $\widetilde{R}_b$:
\begin{equation}\label{eff}
\eta_b\equiv\frac{\mathcal{E}_{\mbox{avail}}-\mathcal{E}_{\mbox{lost}}}{\mathcal{E}_{\mbox{avail}}}=
1-(\widetilde{R}_b/R_b)^4=\frac{\widetilde{Q}_s}{Q_s}
\end{equation}
where $\widetilde{Q}_s$ is the charge of a trapezoidal bunch that is described by Eq. (\ref{trapb}) but terminated at $\xi_s+\Delta\xi_f$ instead of $\xi_s+\frac{r_s^2}{4E_s}$. We note that the efficiency approaches $100\%$ for $\Delta\xi_f\rightarrow \frac{r_s^2}{4E_s}\Rightarrow \widetilde{Q}_s\rightarrow Q_s$. Because the mathematical formulation involves approximations, there is still some energy in the plasma behind the bunch, even with the optimal $\lambda(\xi)$. This is the case in Fig. \ref{figure1}, where a second bubble with radius $R_{b2}\sim R_b/2$ does appear, but because  $Q_s\times E_s\propto R_b^4$, the efficiency is still $\eta_b\sim 90\%$. The wakefield within the bunch in Fig. \ref{figure1} is constant, in agreement with the theory.  We note that  a $10 \%$ deviation of the total charge for a fixed bunch length leads to a wake that is no longer flat.

It is illustrative to compare the amount of charge that can be loaded into linear and nonlinear wakes. 
If we assume for the linear wake 
  an effective $A\simeq c^2/\omega_p^2$,  which is required for high efficiency and good beam quality \cite{Katsouleas}, we have:

\begin{eqnarray}
 \label{linqE}
\frac{Q_l\times E_l}{mc^2/r_e} &=&\frac{1}{8\pi}\times\biggl(\frac{n_1}{n_0}\biggr)^2\times\sin^2(k_p\xi_0)
\\
  \label{prbdn}
\frac{Q_s\times E_s}{mc^2/r_e} &=&\frac{1}{4^3}\times(k_pR_b)^4
\end{eqnarray}
where $r_e=e^2/(mc^2)$ is the classical electron radius. In the linear regime, the density perturbation is $n_1/n_0\ll1$. In the blowout regime, because the total accelerating force scales with the fourth power of the blowout radius, a radius $k_pR_b\sim 5$ leads to a total force $\sim 1000$ times larger than that in the linear regime. 
 Eq. (\ref{prbdn}) can be converted into  an engineering formula:
\begin{equation}
 \label{eng1}
\frac{Q_s}{1nC}\times \frac{eE_s}{mc\omega_p}\simeq 0.047\times\sqrt{\frac{10^{16}cm^{-3}}{n_p}} \times(k_pR_b)^4
\end{equation}

 For a bi-Gaussian beam driver with $k_p\sigma_z\sim1$ and $k_p\sigma_r\ll1$ we have $k_pR_b\simeq 2\sqrt{\frac{n_b}{n_0}(k_p\sigma_r)^2}\equiv 2\sqrt{\Lambda}$  \cite{weipop},  and for a matched laser driver $k_pR_b\simeq 2\sqrt{a_0}$  \cite{weipop}, where $a_0$ is the normalized vector potential. Both for a  beam driver with $3\times10^{10}$ electrons and $\sigma_r\ll\sigma_z=16.8\mu m$  in a plasma with $n_p=10^{17}cm^{-3}$, and for a matched laser-driver with power $P=200TW$ in a plasma with $n_p=1.2\times10^{18}cm^{-3}$, we have $k_pR_b\simeq 4$. Choosing $\frac{eE_s}{mc\omega_p}=\frac{k_pR_b}{2}\simeq 2$, we obtain $Q_s\simeq1.9nC$ for the beam-driven case and $Q_s\simeq0.55nC$ for the laser-driven one.

Another analytically tractable case is that of a bunch with a flat-top profile starting at $\xi=\xi_{\bar{s}}$:   $l(0<r_b\leq  r_{\bar{s}})=l_0$. The behavior of the flat-top bunch is important because it is similar to that of a Gaussian bunch (see below). For such bunches, Eq. (\ref{genwake}) becomes:
\begin{equation}
\label{ezflattop}
E_z(0<r_b\leq r_{\bar{s}})\simeq
-\frac{r_b}{2\sqrt{2}} \sqrt{\frac{8l_0 (r_b^2-r_{\bar{s}}^2) +R_b^4}{r_b^4}-1}
\end{equation}
There are three distinct cases, all of which can be solved analytically. For small charge per unit length $l_0<R_b^4/(8r_{\bar{s}}^2)$,  the plasma electrons reach the $\xi$-axis quickly with some remaining kinetic energy.  For $l_0>R_b^4/(8r_{\bar{s}}^2)$, there is a minimum radius $r_m =\sqrt{4l_0-\sqrt{(4l_0)^2+R_b^4-8l_0r_{\bar{s}}^2}}$, for which $E_z(\xi_m)=0$ and the transverse velocity of the  innermost particle changes sign as described earlier. At $\xi_m$, the bunch must be terminated; otherwise, for $\xi>\xi_m$, it will experience a decelerating field. Because the plasma electrons do not return on the $\xi$-axis, they still have some potential energy. Thus, for $l_0\neq R_b^4/(8r_{\bar{s}}^2)$, there is  always some energy in the plasma behind the bunch.

For a flat-top bunch, the beam-loading efficiency is maximized  if $l_0=R_b^4/(8r_{\bar{s}}^2)$.  Then the shape of the bubble and the wakefield are given by:  
\begin{eqnarray}\label{elipse}
  8l_0 &=&r_b^2+\frac{1}{2}\biggl(\xi- \xi_{\bar{s}}+\sqrt{2}\sqrt{8l_0-r_{\bar{s}}^2}\biggr)^2
\\
 \label{slope14}
   E_z&=&   -\frac{1}{4}(\xi-\xi_{\bar{s}}) + E_z(\xi=\xi_{\bar{s}})
 \end{eqnarray}
and the innermost particle will reach the $\xi$-axis at $\xi_{\bar{s}}+\Delta\xi_{\bar{s}}$, where $
\Delta\xi_{\bar{s}}=\frac{\sqrt{2}}{r_{\bar{s}}}(R_b^2-\sqrt{R_b^4-r_{\bar{s}}^4})$. In this case, the energy absorption per unit length is identical to that of an optimal trapezoidal bunch $2\pi l_0\Delta\xi_{\bar{s}}\langle  \vert E_z\vert \rangle=Q_s\times E_s$.
The difference in the accelerating force experienced by the front and the back of the bunch will  tend to increase the bunch's energy spread. This can be avoided  either by injecting the bunch with an initial energy chirp  to compensate for the effect caused by the field in Eq. (\ref{slope14}) or by using a monoenergetic trapezoidal bunch.

If the driver travels with a velocity slower than that of the accelerating electrons, these 
 electrons will move with respect to the wake. 
 In this context, it is interesting to see what happens if a flat-top electron bunch optimized for some $\xi_1$ is 
  instead placed at  $\xi_2$ and $\xi_3$, both smaller than $\xi_1$.

\begin{figure}[!thb]
\begin{center}
\includegraphics[width=3.2 in]{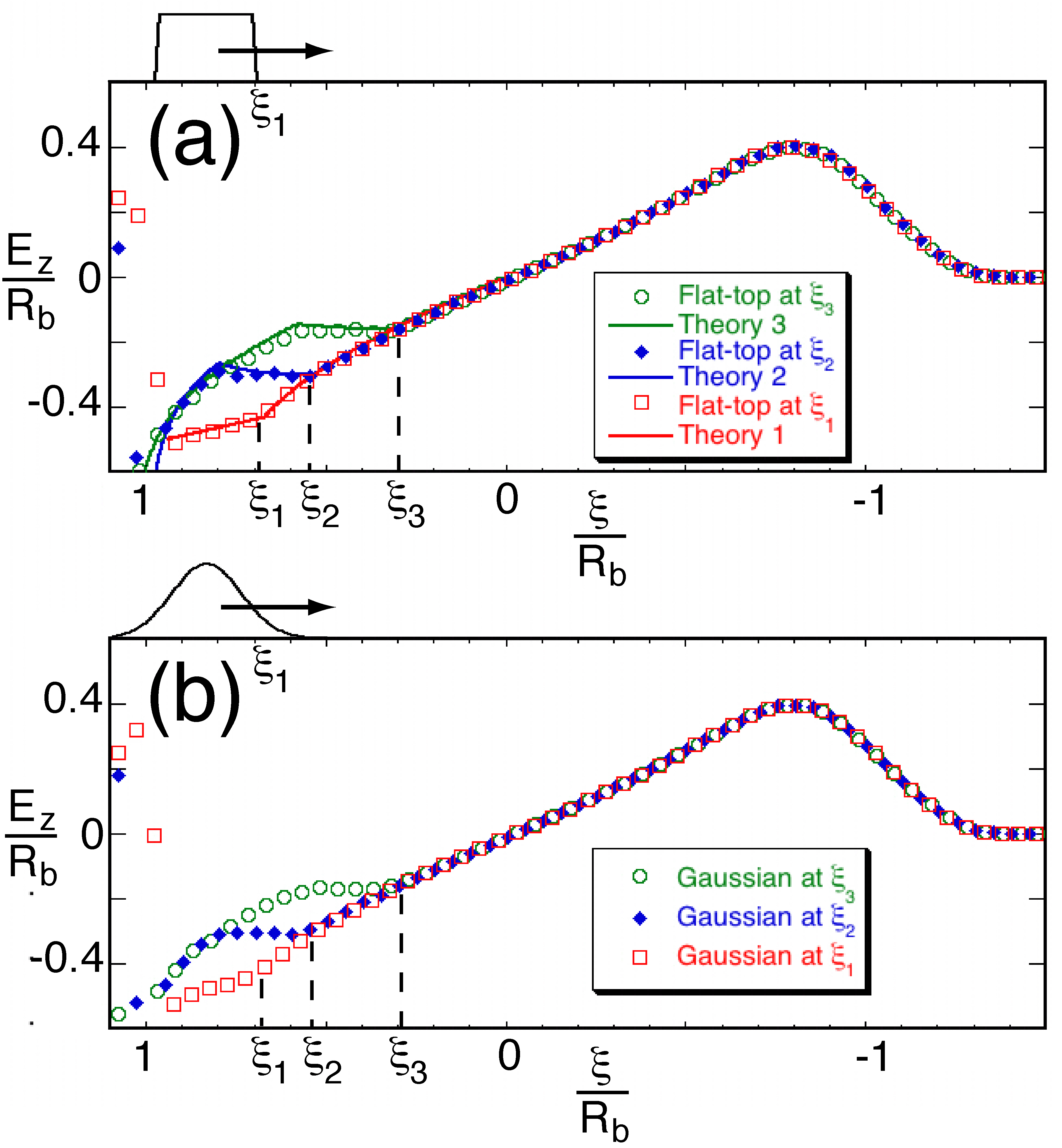}
\caption{Wakefield lineouts for (a)  
 a flat-top electron bunch and (b) a Gaussian bunch with the same charge at three different locations $\xi_1\mbox{(red)},\xi_2\mbox{(blue)},\xi_3\mbox{(green)}$ is plotted from theory (solid lines (a)) and simulations (symbols (a), (b)).} 
\label{figure2}
\end{center}
\end{figure}

In Fig. \ref{figure2}(a), 
we compare the lineouts of the wakefield $E_z(\xi,r_b=0)$ from three 2D cylindrically symmetric  simulations with the theoretical results for flat-top beams.  
  For each simulation, an 
 electron bunch with $l_0=0.25R_b^2$ and length $\Delta\xi_{\bar{s}}=0.27R_b$ is loaded  at one of  three locations: $\xi_1=0.67R_b, \xi_2=0.53R_b, \xi_3=0.31R_b$. 
The open red squares correspond to loading at $\xi_1$, the solid blue  
diamonds to $\xi_2$, and the open green circles to $\xi_3$.   The solid lines are derived from the theory  (for $l_0>R_b^4/(8r_{\bar{s}}^2)$,  the particle trajectory in the region $\xi_{\bar{s}}\leq\xi<\xi_m$ can be written in terms of the integral $E(\varphi\vert m)$) and are in excellent agreement with the simulations in all three cases.

We repeated the simulations using Gaussian bunches with the same number of particles as in the flat-top cases 
and $N_b(z)=\frac{N_b}{\sqrt{2\pi}\sigma_z}e^{-z^2/(2\sigma_z^2)}$, where $\sigma_z= \Delta\xi_{\bar{s}}/(2\sqrt{2})$. Each bunch is placed so that its center is at a distance $\sqrt{2}\sigma_z$  from $\xi_1,\xi_2$, and $\xi_3$ for the three simulations. The results, shown in Fig. \ref{figure2}(b), confirm that the Gaussian bunches may be treated using the theory for flat-top bunches. In both Fig. \ref{figure2}(a) and \ref{figure2}(b), we observe that the wakefield is relatively flat regardless of the placement of the bunch.  The initial negative slope is balanced by a smaller positive slope for most of the acceleration process.

 The main approximation in the theory, $k_pR_b\gg1$, 
 is expected to hold for $k_pR_b\gtrsim3$. However, the formalism described here can still be applied if one numerically solves Eq. ($11$) of Ref. \cite{WEI}.  
 
Work supported by the Department of Energy under grants DE-FG02-03ER54721,  DE-FG03-92ER40727, DE-FG52-06NA26195, and  DE-FC02-07ER41500. Simulations were carried out on the DAWSON Cluster funded under an NSF grant, NSF-Phy-0321345, and at NERSC.

\end{document}